\documentclass{ckm}                 
\usepackage{txfonts}            

\confname{Workshop on the CKM Unitarity Triangle, IPPP Durham, April
  2003}

\usepackage[dvips]{color}
\definecolor{darkgreen}{rgb}{0, 0.85, 0}
\definecolor{darkyellow}{rgb}{0.85, 0.85, 0}

\newcommand{\un}[2]{\ensuremath{\mathrm{#1 \, #2}}}

\newcommand{\units}[1]{\ensuremath{\mathrm{#1}}}

\newcommand{\degrees}[1]{\ensuremath{\mathrm{#1^{\circ}}}}

\newcommand{\No}{\ensuremath{\mathrm{N^{\underline{o}}}}}

\newcommand{\E}[1]{\ensuremath{\cdot 10^{#1}}}

\newcommand{\prt}[1]{\ensuremath{{\rm #1}}}
\newcommand{\qrk}[1]{\ensuremath{#1}}
\newcommand{\Bp}{\prt{B^+}}
\newcommand{\Bm}{\prt{B^-}}

\newcommand{\Bso}{\prt{B_{s}^0}}
\newcommand{\Bsob}{\prt{\overline{B_{s}^0}}}
\newcommand{\Bdo}{\prt{B_{d}^0}}
\newcommand{\Bdob}{\prt{\overline{B_{d}^0}}}

\newcommand{\bbbar}{\qrk{b\overline{b}}}

\newcommand{\DZERO}{{D\O}}
\newcommand{\Gs}{\ensuremath{\Gamma_s}}
\newcommand{\DGs}{\ensuremath{\Delta\Gs}}
\newcommand{\DGsR}{\ensuremath{\frac{\DGs}{\Gs}}}

\newcommand{\Gd}{\ensuremath{\Gamma_d}}
\newcommand{\DGd}{\ensuremath{\Delta\Gd}}
\newcommand{\DGdR}{\ensuremath{\frac{\DGd}{\Gd}}}

\newcommand{\babar}{\textsc{BaBar}}
\newcommand{\belle}{\textsc{Belle}}

\newcommand{\fig}{.}

\title{Heavy Flavour Lifetimes and Lifetime Differences}

\author{Jonas Rademacker\addressmark{a}}

\address[a]{CDF Experiment, Fermilab, and University of Oxford}

\begin{document}

\begin{abstract}
We give an overview of heavy flavour lifetime measurements, focusing
on recent results from the Tevatron and the B factories.
\end{abstract}

\maketitle


\section{Introduction}

 In the first part of this article we summarise the status and latest
 measurements of B-hadron lifetimes and lifetime ratios, including
 some recent result from the Tevatron and the B factories, and
 compare those results with the predictions from Heavy Quark Expansion
 (HQE). Future prospects for lifetime measurements at the B factories
 and the Tevatron are discussed.

 In the second part, we review the status and prospects of measuring
 the difference between the lifetimes of the two CP eigenstates in the
 \Bso-\Bsob\ system.

\section{Lifetimes and Lifetime Ratios}
\subsection{Theoretical Predictions on B Hadron Lifetimes}
 Life time measurements in the heavy quark sector gain specific
 significance due to the precise predictions of Heavy Quark expansion
 (see e.g.~\cite{Uraltsev:1998bk},~\cite{Shifman:2000jv}) 
 thus providing a testing ground for this theoretical tool that is
 frequently used, for example to relate experimental measurements to
 CKM parameters like $\Gamma_d$ to $\left|V_{cb}\right|$ or $\Delta
 m_s/\Delta m_d$ to $\left|V_{ts}/V_{td}\right|$.

 The hierachy expected for b hadron lifetimes is~\cite{BPhysicsAtTev}:
\begin{eqnarray*}
 & & \tau(\prt{B_c}) \ll \tau(\Xi_b^0) \\
 & & \color{red} \sim \tau(\prt{\Lambda_b}) < \tau(\prt{B_d^0}) \sim \tau(\prt{B_s^0})< \tau(\prt{B^-}) \\
 & & < \tau(\prt{\Xi_b^-}) < \tau(\prt{\Omega_b}).
\end{eqnarray*}\\
 Recent HQE predictions for the lifetime ratios are~\cite{Franco:2002fc}:

\begin{itemize}
 \setlength{\itemsep}{2ex plus2ex minus2ex}
 \item $\tau(\prt{B^-})/\tau(\prt{B_d^0})       = 1.06 \pm 0.02$
 \item $\tau(\prt{B_s})/\tau(\prt{B_d^0})       = 1.00 \pm 0.01$
 \item $\tau(\prt{\Lambda_b})/\tau(\prt{B_d^0}) = 0.90 \pm 0.05$
\end{itemize}

\subsection{The B Factories}

 The B factories \babar\ and \belle\ at the asymmetric \prt{e^+ e^-}
 colliders PEP-II and KEK have collected \un{123}{fb^{-1}} and
 \un{145}{fb^{-1}} worth of data respectively up to May 2003,
 running at the \prt{\Upsilon(4S)} resonance. 

\subsubsection{Lifetimes at the B Factories: Method}
 The \prt{\Upsilon(4S)} decays to \Bdo, \Bdob\ or \Bp, \Bm, nearly at
 rest in the \prt{\Upsilon(4S)} rest frame.  By colliding \prt{e^+}
 and \prt{e^-} of different energies, the CM frame is boosted
 ($\beta\gamma \sim 0.54$ for \babar\ and $\sim 0.43$ for \belle) such
 that the \Bdo\ and \Bdob\ travel a measurable distance in the
 detector before decaying.

 Because the B mesons decay virtually at rest in the
 \prt{\Upsilon(4S)} frame, their momenta in the lab frame are known
 from the beam momentum. This constrains the decay dynamics
 considerably with the important consequence that the decay vertex of
 a B meson can be obtained from a single decay product, by
 intersecting its track with the beam axis. The decay distance along
 the beamline ($z$) is directly proportional to the proper decay time
 for a given beam momentum (small corrections
 apply~\cite{babarHadronic2000Conf}). Lacking primary vertex
 information, it is the distance between the decay vertices of the two
 B mesons that is used for measuring the life time. This distance is
 typically \un{\sim 250}{\mu m} and \un{\sim 200}{\mu m} at \babar\
 and \belle\ respectively.

 In the standard method, one B meson is fully reconstructed,
 (\prt{B_{rec}}), and another one partially, from as little as one or
 two tracks (\prt{B_{opp}}), with a correspondingly degraded vertex
 resolution. The lifetime difference is calculated from the
 difference in the position along the beam line ($\Delta z=
 z_{\mathrm{B_{reco}}} - z_{\mathrm{opp}}$) of the two B vertices.
\begin{figure}
\hbox to\hsize{\hss
\includegraphics[width=\hsize]{%
\fig/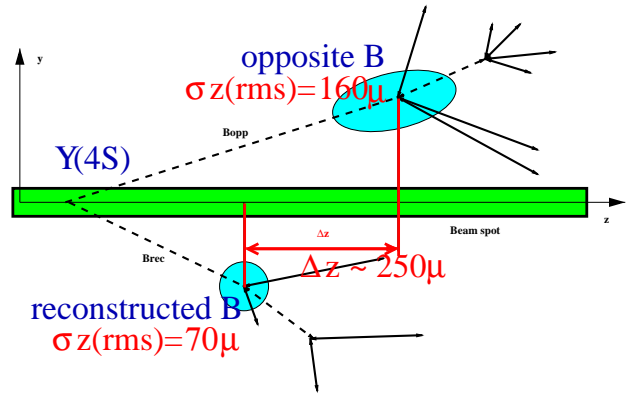}
\hss}
\caption{Schematic of lifetime measurements at \babar\
(from~\cite{babarHadronic2000Conf} with added comments).}
\label{fig:babarDiagram}
\end{figure}
 This is illustrated in Fig.~\ref{fig:babarDiagram}. The resolution
 function is modeled using Monte Carlo simulation.
\begin{figure}
\hbox to\hsize{\hss
\includegraphics[width=\hsize]{\fig/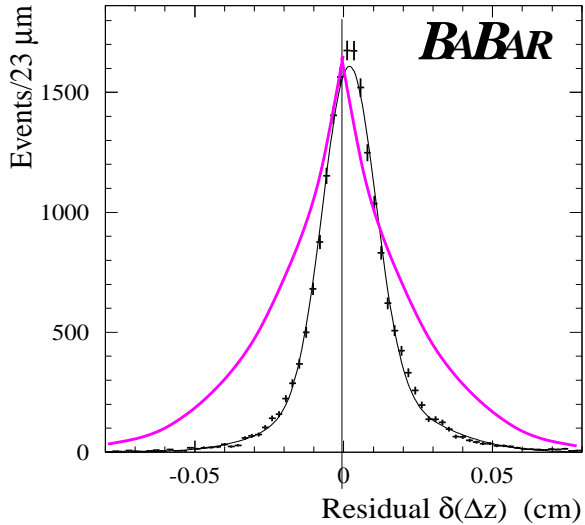}
\hss}
\caption{$\Delta z$ Resolution at \babar\ for \prt{B^+\to J/\psi K^+}
 \cite{babarHadronic2000Conf}, with $\exp(\Delta z/\un{250}{\mu m})$ 
 superimposed for illustration.}
\label{fig:babarResolution}
\end{figure}
 In Fig.~\ref{fig:babarResolution}, the Monte Carlo generated
 resolution function for $\Delta z$ at \babar\ for the decay
 \prt{B^+\to J/\psi K^+} is shown~\cite{babarHadronic2000Conf}. An
 exponential with a mean decay distance of \un{250}{mu m},
 representing approximately the signal distribution before detector
 effects, is superimposed, illustrating how the signal is of a similar
 width as the resolution function, which must therefore be modeled
 carefully. This modeling of the resolution function, together with
 the modeling of the background distribution, is the biggest
 systematic uncertainty in both experiments. The so-called
 ``outliers'', a relatively small number of events with very large
 reconstructed $\Delta z$, represent a particular problem. Both
 experiments are able to control it well enough to keep the systematic
 error below the statistical uncertainty.

 Both experiments describe the $\Delta t = \Delta z/\left(c
 \left(\beta\gamma\right)_{\Upsilon}\right)$ distribution in terms of
 three components: signal, background and outliers.  The beam
 constrained mass
\begin{figure}
\hbox to\hsize{\hss
\includegraphics[width=\hsize]{\fig/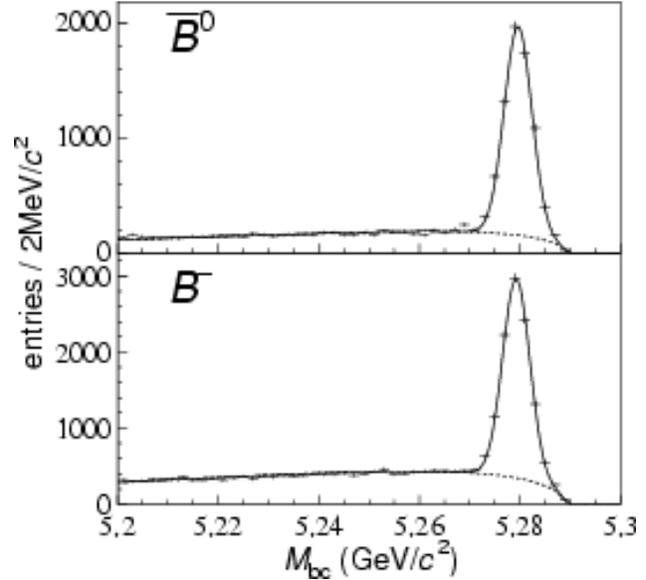}
\hss}
\caption{\No\ of events vs Beam Constrained Mass at \belle\ \cite{belleHadronic2002}}
\label{fig:belleMass}
\end{figure}
 (shown in Fig.~\ref{fig:belleMass} for \belle) is used for an
 event-by-event signal probability. The fraction of outliers is a free
 fit parameter. The \Bdo\ and \Bp\ distributions are fit
 simultaneously in an unbinned likelihood fit. Besides these
 commonalities, there are some differences in the event selection
 and modeling of the resolution function which are described in
 detail in the publications by the respective experiments
 \cite{babarHadronic2001} \cite{belleHadronic2002}.

\subsubsection{Results}
\paragraph{\belle}
 Using the following fully reconstructed hadronic decays:
 \prt{B^0 \to D^{(*)-}(\pi^+, \rho^+), J/\psi
 K^0_S, J/\psi K^{*0}}, \prt{B^+ \to \overline{D}^0 \pi^+, J/\psi
 K^+}, \belle\ find the following \Bdo\ and \Bp\ lifetimes~\cite{belleHadronic2002}:
\begin{eqnarray*}
 \tau_{\prt{B_d^0}}                  &=& \un{1.554\pm 0.030 \pm 0.019}{ps}\\
 \tau_{\prt{B^+}}                    &=& \un{1.695\pm 0.026 \pm 0.015}{ps}\\
 \tau_{\prt{B^+}}/\tau_{\prt{B_d^0}} &=& \un{1.091\pm 0.023 \pm 0.014}{}
\end{eqnarray*}\\
\begin{figure}
\hbox to\hsize{\hss
\rotatebox{270}{\includegraphics[height=\hsize]{\fig/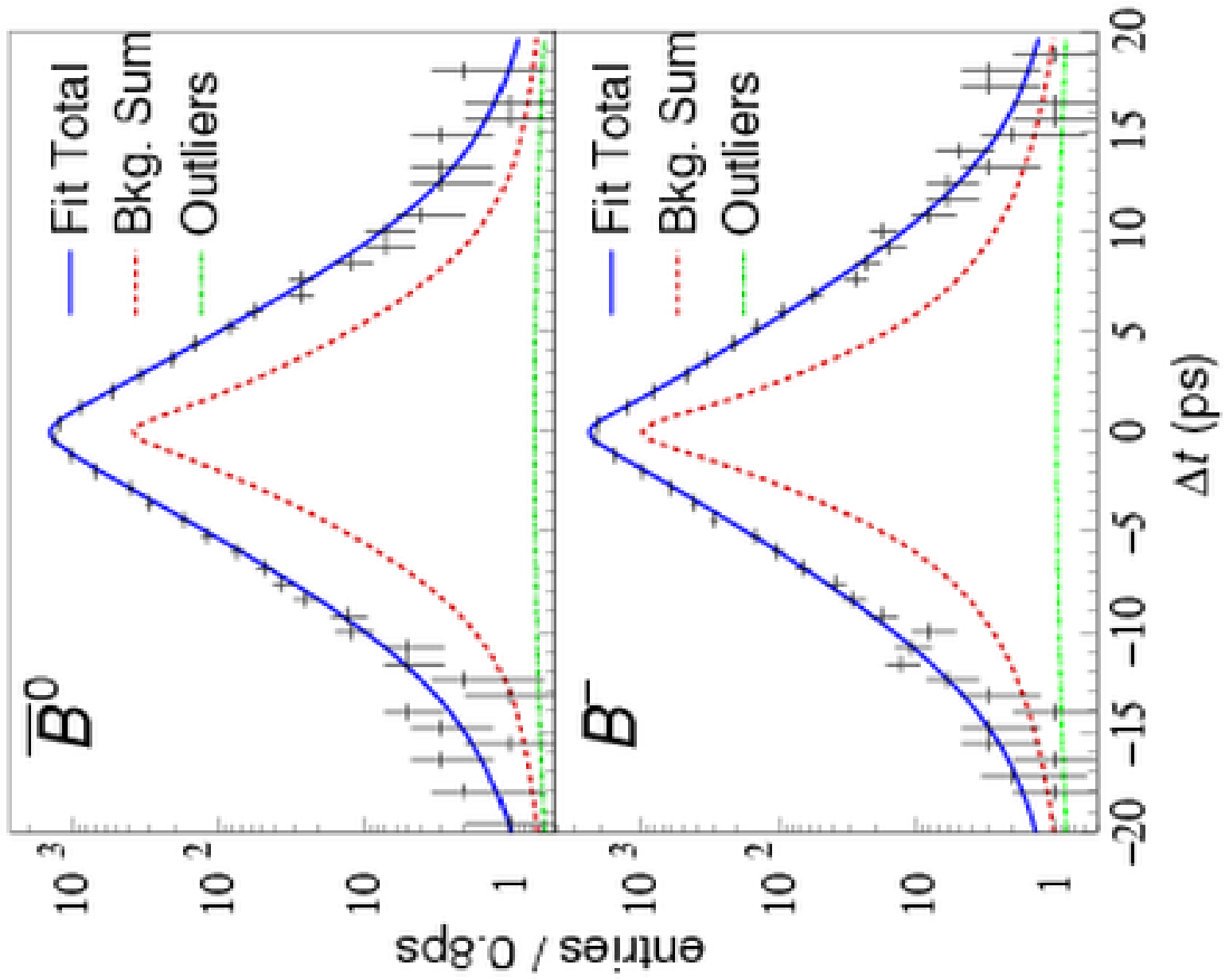}}
\hss}
\caption{\belle's life time fit~\cite{belleHadronic2002}.}
\label{fig:belleFit}
\end{figure}
The fit result to the data, showing separately the background and the
outlier contribution, is shown in Fig.~\ref{fig:belleFit}.

\paragraph{\babar\ Fully Hadronic}
 Using the following fully reconstructed hadronic decays:
 \prt{B^0 \to D^{(*)-}(\pi^+, \rho^+, a_1^+), J/\psi
   K^0_S, J/\psi K^{*0}} and 
   \prt{B^+ \to \overline{D}^0 \pi^+, J/\psi K^+, \psi(2S) K^+}
 \babar\ find the following \Bdo\ and \Bp\
 lifetimes~\cite{babarHadronic2001}:
\begin{eqnarray*}
\tau_{\prt{B_d^0}}                   &=& \un{1.546\pm 0.032 \pm 0.022}{ps}\\
\tau_{\prt{B^+}}                     &=& \un{1.673\pm 0.032 \pm 0.023}{ps} \\
\tau_{\prt{B^+}}/\tau_{\prt{B_d^0}}  &=& \un{1.082\pm 0.026 \pm 0.012}{}
\end{eqnarray*}\\
\begin{figure}
\hbox to\hsize{\hss
\includegraphics[width=\hsize]{\fig/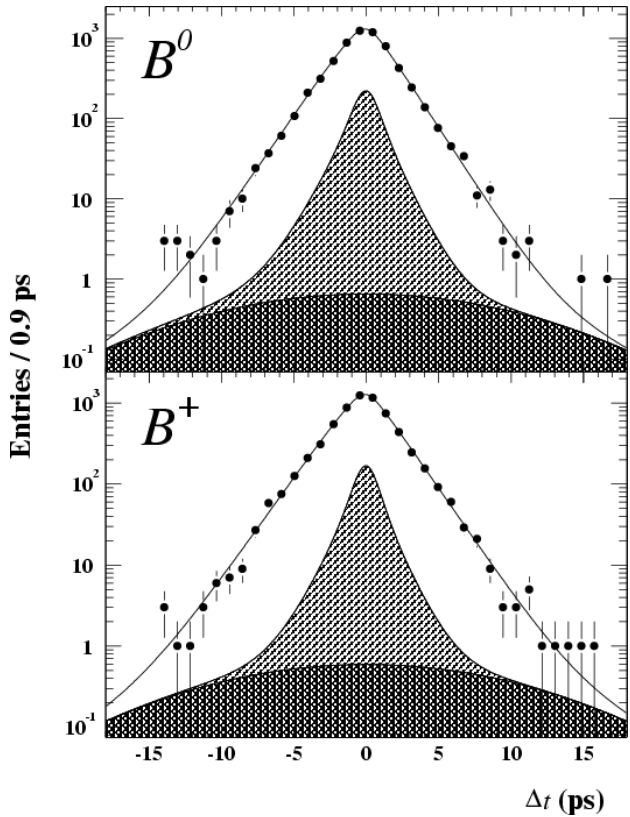}
\hss}
\caption[\babar's life time fit]{\babar's life time fit
\cite{babarHadronic2001}. 
Solid line: total fit. Single hatched: total background. Cross hatched: outliers.}
\label{fig:babarFit}
\end{figure}
The fit result to the data is shown in Fig.~\ref{fig:babarFit}.

\paragraph{More results from \babar}
As mentioned earlier, the decay kinematics at \babar\ and \belle\
allow to find the $z$ position of a decay vertex from as little as a
single track. While in the previously mentioned measurements, one of
the pair of B's is fully reconstructed, \babar\ also published a set
of measurements where also the \prt{B_{\mathrm{rec}}} is reconstructed
partially.  These are summarised in Table~\ref{tab:babarPartial}.
\begin{table}
\begin{center}
\begin{tabular}{||r|c|}
\hline\hline
&
\prt{B^0\to D^{*-}\mbox(partial) \ell^+ \nu} \hfill \cite{babarPartialDstLep2002}
\\\hline 
$\tau_{\prt{B_d^0}}$                  
& \un{1.529 \pm 0.012 \pm 0.029}{ps}
\\\hline\hline
&
 \prt{B^0\to D^{*-}\mbox(partial)(\pi^+, \rho^+)} \hfill \cite{babarPartialDstPi2002}
\\\hline
$\tau_{\prt{B_d^0}}$                  
& 
 \un{1.533 \pm 0.034 \pm 0.038}{ps}
\\\hline\hline
&
\prt{B^0\to D^{*-} \ell^+ \nu} \hfill \cite{babarFullDstLep2002}
\\\hline 
$\tau_{\prt{B_d^0}}$                  
& \un{1.523^{+0.024}_{-0.023} \pm 0.022}{ps}
\\\hline\hline
&
Di-lepton (prelim) \hfill \cite{babarDilepton2002}
\\\hline 
$\tau_{\prt{B_d^0}}$                  
& \un{1.557\pm 0.028 \pm 0.027}{ps}
\\
$\tau_{\prt{B^+}}$                    
& \un{1.655\pm 0.026 \pm 0.027}{ps}
\\
$\tau_{\prt{B^+}}/\tau_{\prt{B_d^0}}$ 
&\un{1.064\pm 0.031 \pm 0.026}{ps}
\\\hline\hline
\end{tabular}
\end{center}
\caption{\babar's results from partially reconstructed decays. Here, a
``partial \prt{D^*}'' is a \prt{D^*} decaying to to \prt{D^0 \pi},
reconstructed using kinematic constraints and the pion momentum, only,
without reconstructing the \prt{D^0}~\cite{babarPartialDstPi2002}.}
\label{tab:babarPartial}
\end{table}

\subsubsection{Status of Lifetime Measurements Including Results from \babar\ and \belle}
\begin{figure}
\hbox to\hsize{\hss
\includegraphics[width=.7\hsize]{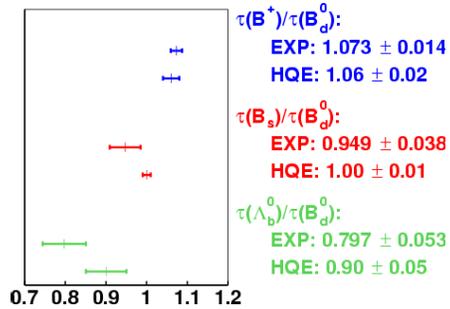}
\hss}
\caption{Status of lifetime ratio measurements, incl. results
from \babar, \belle, CDF, LEP. $\tau(\prt{B_s})/\tau(\prt{B_d})$ 
and $\tau(\prt{B^+})/\tau(\prt{B_d})$ from
\cite{heavyFlavourAveraging}, $\tau(\prt{\Lambda_b})/\tau(\prt{B_d})$ 
from \cite{pdg2002}.}
\label{fig:bfactoriesConclusion}
\end{figure}
 Since the B factories have started taking data, they have reduced the
 error on $\tau_{\prt{B^+}}/\tau_{\prt{B_d^0}}$ by half.
 Fig.~\ref{fig:bfactoriesConclusion} shows the current status of the
 life time measurements and compares them with HQE predictions. The
 $\tau_{\prt{B^+}}/\tau_{\prt{B_d^0}}$ measurement, dominated by the
 precise results from the B factories, is already more precise than
 that of the HQE prediction, and we can expect further improvements in
 the near future.

 The situation is different for the \prt{B_s} and the $\Lambda_b$,
 which are not accessible at the B factories. The experimental
 precision of the $\tau_{\prt{B_s^0}}/\tau_{\prt{B_d^0}}$ measurement
 lags behind that of the HQE calculations. For the $\Lambda_b$,
 experiment and theory don't appear to be in very good agreement, but
 the experimental and theoretical uncertainties are still rather
 large. Improved measurements and calculations are needed for
 clarification.

 Both, \prt{B_s} and \prt{\Lambda_b} particles are produced abundantly
 at hadron colliders, from where we expect dramatically improved
 lifetime measurements in the near future. The hadron collider
 currently producing large numbers of \prt{B_s} and \prt{\Lambda_b} is
 the Tevatron at Fermilab.

\subsection{Lifetimes at the Tevatron}
\subsubsection{Run~II}
 CDF and \DZERO\ have been taking data at Tevatron Run~IIa for
 about two years. For \prt{p\bar{p}} collisions at \un{1.96}{TeV}, the
 \bbbar\ production cross section is $\sigma_{\bbbar} \sim
 \un{0.1}{mb}$.
\begin{figure}
\hbox to\hsize{\hss
\includegraphics[width=\hsize]{\fig/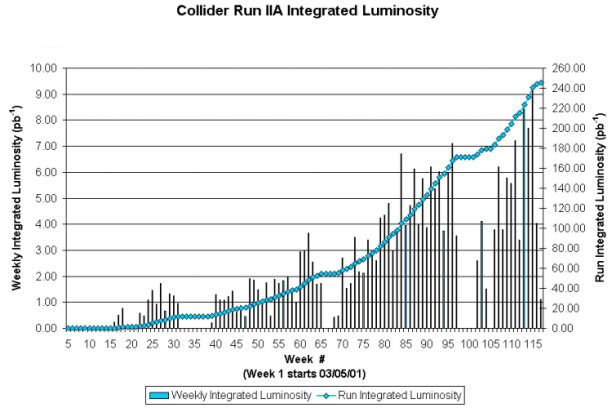}
\hss}
\caption{Luminosity at the Tevatron}
\label{fig:TevLumi}
\end{figure}
 The integrated luminosity delivered until June~2003 is shown in
 Fig.~\ref{fig:TevLumi}. The integrated luminosity at Run~IIa is
 expected to be \un{2}{fb^{-1}}.
\begin{table}
\begin{center}
Projected $\int\mathcal{L} dt\;/\; (\units{fb^{-1}})$\\
\begin{tabular}{|l|r@{.}l|r@{.}l|}
\hline
Year & 
\multicolumn{2}{c|}{Baseline} &  
\multicolumn{2}{c|}{Stretch} \\\hline
2002 & 0&08 & 0&08 \\\hline
2003 & 0&2  & 0&32 \\\hline
2004 & 0&4  & 0&6  \\\hline
2005 & 1&0  & 1&5  \\\hline
2006 & 1&5  & 2&5  \\\hline
2007 & 1&5  & 3&0  \\\hline
2008 & 1&8  & 3&0  \\\hline
\hline
Total & \hspace{1em}6&5 & \hspace{.5em}11& \\\hline
\end{tabular}
\end{center}
\caption{Projected integrated luminosity at the Tevatron for baseline and
best-case (``stretch'') scenario.  The total
integrated luminosity by 2008 is expected to be between
\un{6}{fb^{-1}} and \un{11}{fb^{-1}}.}
\label{tab:TevLumiProj}
\end{table}

 The projected luminosity for each year until 2008 is listed in
 Table~\ref{tab:TevLumiProj}, for two scenarios: The base-line
 scenario, and a best-case scenario (``stretch''). The total
 integrated luminosity at the end of Run~II in 2008 is expected to lie
 between \un{6}{fb^{-1}} and \un{11}{fb^{-1}}.

\subsubsection{\DZERO\ and CDF}
 Both experiments at the Tevatron have undergone major upgrades for
 Run~II, optimising their B~physics potential. The most significant
 upgrade at \DZERO\ is the introduction of a magnetic field and a new
 tracking system providing precise momentum information. This significantly
 improves the mass resolution. CDF also improved its tracking with a new,
 faster drift chamber. Both experiments have new Silicon vertex
 trackers providing excellent proper time resolution, sufficient to
 resolve the expected fast oscillations in the \Bso\ system. The
 excellent impact parameter resolution is used for triggering on
 B-events. Both experiments have increased their muon coverage since
 Run~I, and have an efficient di-muon trigger for finding \prt{B^0 \to
 J/\psi X} decays. \DZERO's $\mu-$trigger covers a particularly large
 pseudo rapidity range up to $\left| \eta \right| = 2$.

\paragraph{IP Trigger}
 One of the most innovative improvements for B physics at the Tevatron
 is the large-bandwidth hadron trigger at CDF, which triggers on the
 impact parameters of tracks at Level~2. The eXtremely Fast Tracker
 (XFT) uses pattern matching to find tracks in the COT (drift chamber)
 within \un{5.5}{\mu s}, with about $96\%$ efficiency for momenta
 above \un{1.5}{GeV}. These XFT tracks are combined with tracks in the
 Silicon Vertex Detector by the Silicon Vertex Tracker (SVT), which
 makes impact parameter information available at Level~2 to a
 precision of \un{\sim 50}{\mu m}. The 2-Track hadron trigger combines
 the information on the direction (XFT), momentum (XFT) and impact
 parameter (SVT) to trigger on hadronic B decays.
\begin{table}
\fbox{\parbox{0.95\columnwidth}{
  {L1:} 2 XFT tracks, $p_t > \un{2}{GeV}$, $\Delta\phi < \degrees{135}$, $p_{t1} + p_{t2} > \un{5.5}{GeV}$.\vspace{1ex}\\
  {L2:}\\
  \begin{tabular}{p{0.43\columnwidth}|p{0.43\columnwidth}}
   2-body:\newline e.g. \prt{\Bdo \to \pi\pi} 
                      & Multi-body:\newline e.g. \prt{\Bso \to D_s \pi} 
   \\\hline
  $\un{100}{\mu m}<IP<\un{1}{mm}$ & $\un{120}{\mu m}<IP<\un{1}{mm}$        \\
   $\degrees{20} < \Delta\phi < \degrees{135}$ 
                      & $\degrees{2} < \Delta\phi < \degrees{90}$ \\
   $L_{xy}$\un{>200}{\mu m}
                      & $L_{xy}$\un{>200}{\mu m}         \\
   IP of B \un{<140}{\mu m} & --                            \\
  \end{tabular}\vspace{1ex}\\
  {L3:} Same with refined tracks \& mass cuts.
}}
\caption[The CDF hadron trigger]{The CDF hadron trigger. $\Delta\phi$ is the angle between the
 tracks in the transverse plane. IP is the impact parameter in that
 plane. $\mathrm{L_{xy}}$ is the decay length in the transverse plane,
 which can be calculated from the impact parameters and $\Delta\phi$.}
 \label{tab:hadronTriggerCDF}
\end{table}
 The trigger requirements for the two scenarios, 2-body and multi-body
 B decays, are given in Table~\ref{tab:hadronTriggerCDF}. The
 SVT+lepton trigger for semileptonic B decays has impact parameter
 requirements on one track only and requires additionally an electron
 or muon with $p_t > \un{4}{GeV}$. 

 \DZERO\ also has impact parameter information available at Level~2, and
 will have a lepton+displaced track trigger, which was however not
 yet available for the data presented here.

 For lifetime measurements it is essential that the bias due to the
 impact parameter cuts in the trigger is corrected for. We will first
 consider measurements that do not suffer from such a trigger bias,
 and then those that do.

\subsubsection{Measurements Using Fully Reconstructed Decays, Without
		IP Trigger}
 Both experiments have published results from fully reconstructed
 hadronic \prt{B\to J/\psi X} decays from the dimuon trigger, which
 are not biased by any impact parameter cut.
\begin{figure}
\hbox to\hsize{\hss
\includegraphics[width=\hsize]{\fig/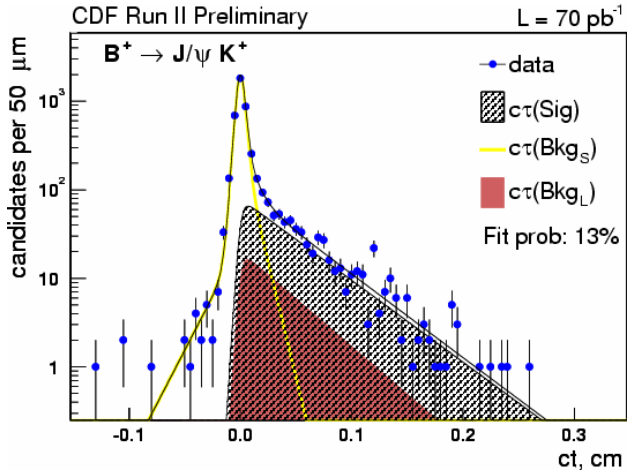}
\hss}
\caption{Projection of Fit to $c\tau_{B^+}$ from \prt{B_u^+ \to
J/\psi(\mu^+\mu^-) K^+} at CDF, using \un{70}{pb^{-1}} of data.}
\label{fig:cdfBplusJPsiFit}
\end{figure}
 An example fit (CDF, \un{70}{pb^{-1}},\prt{B_u^+ \to
 J/\psi(\mu^+\mu^-) K^+}) is shown in
 Fig.~\ref{fig:cdfBplusJPsiFit}. The signal is modeled with an
 exponential, the background by a prompt component and two positive
 and one negative exponential tails (only one positive tail for \Bso\
 because of lower statistics). Signal and background function are
 convolved with a single Gaussian to take into account detector
 effects. The B mass is fit simultaneously and provides an
 event-by-event signal probability.
\begin{table}
\begin{center}
\fbox{\parbox{\hsize}{
Absolute Lifetimes (\DZERO\ \cite{dzeroBJpsiExclusive} and CDF
\cite{cdfBJpsiExclusive}, Run~II prelim.)\\
\begin{tabular}{||lcl||}
\hline\hline
\prt{B_u^+ \to J/\psi(\mu^+\mu^-) K^+} & \DZERO\ & 
                    \un{1.76\pm 0.24(stat)}{ps} \\
\prt{B_u^+ \to J/\psi(\mu^+\mu^-) K^+} & CDF & 
                    \un{1.57\pm 0.07 \pm 0.02}{ps} \\\hline
\prt{B_d^0 \to J/\psi(\mu^+\mu^-)K^{*0}} & CDF & 
                    \un{1.42 \pm0.09\pm 0.02}{ps}\\\hline
\prt{B_s^0 \to J/\psi(\mu^+\mu^-)\phi} & CDF &
                    \un{1.26\pm 0.20\pm 0.02}{ps}
\\\hline\hline
\end{tabular}\vspace{1ex}\\
Lifetime Ratios (CDF~\cite{cdfBJpsiExclusive}, Run~II prelim.,
compared with world average (ave)~\cite{heavyFlavourAveraging} and HQE
predictions~\cite{CKMyellow2002}):\\
{
\includegraphics[width=0.95\columnwidth]{%
\fig/lifeTimeRatiosCDFII_noLambdaProc.epsi}
}
}}
\caption{Lifetimes from \prt{B \to J/\psi X} at Tevatron Run~II. CDF
data correspond to an integrated luminosity of \un{70}{pb^{-1}}.}
\end{center}
\label{tab:BJPsiFull}
\end{table}
\begin{figure}
\begin{tabular}{cc}
{Sidebands} &
{Mass (\un{L_{xy}>400}{\mu m})} \\
\parbox[t]{0.35\columnwidth}{
left sideband:\\
\includegraphics[width=0.35\columnwidth]{%
\fig/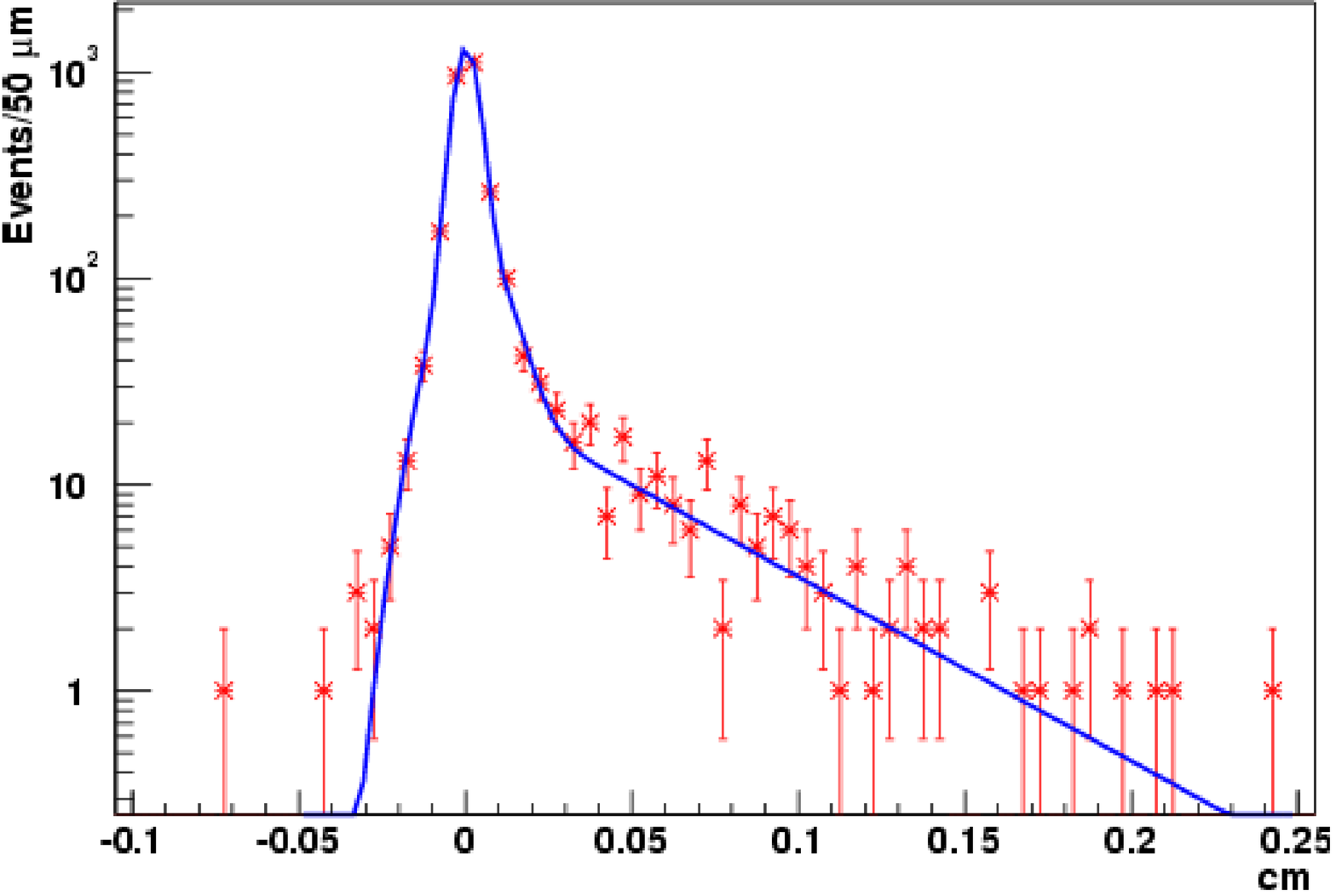}\\
right sideband\\
\includegraphics[width=0.35\columnwidth]{%
\fig/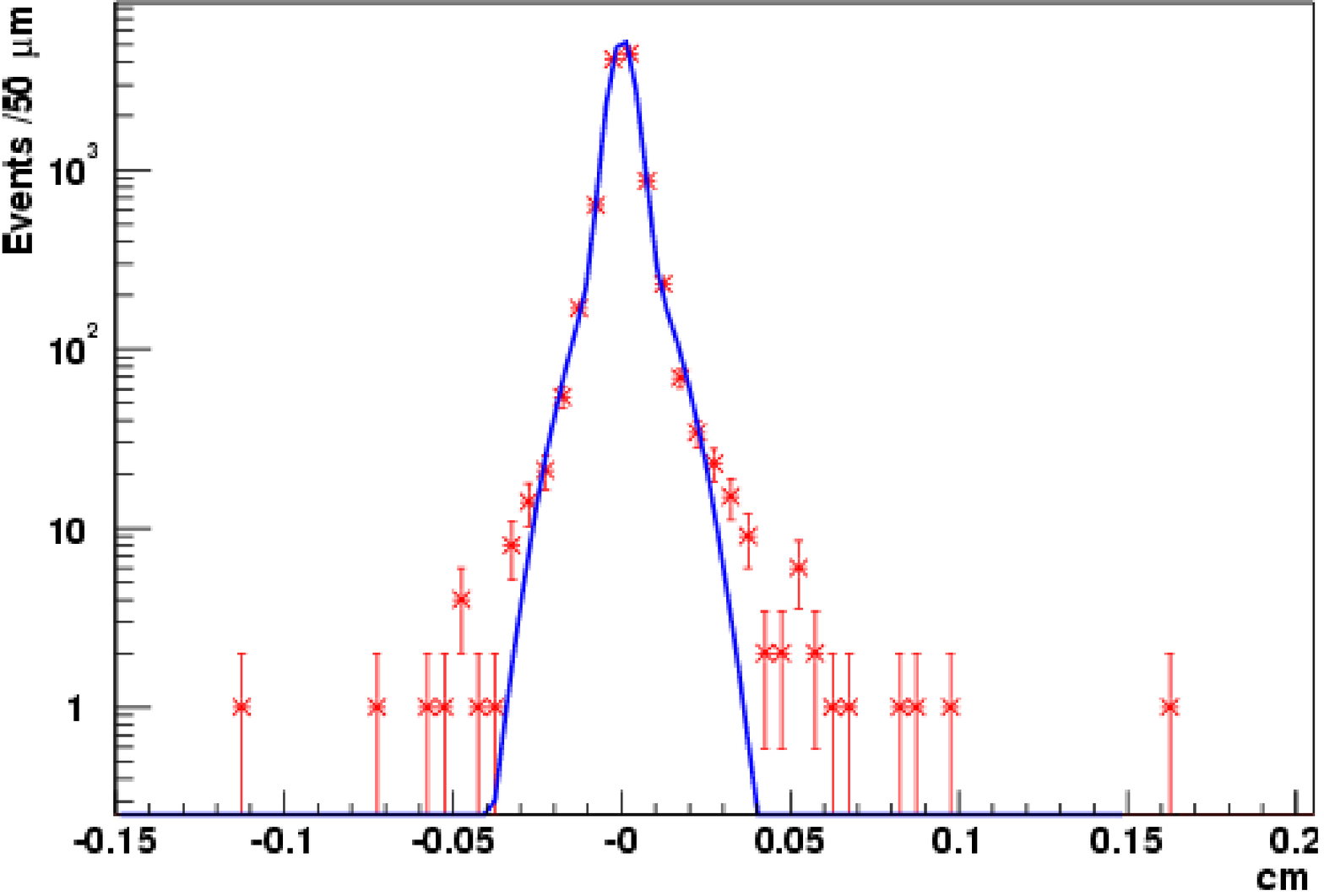}
}
&
\parbox[t]{0.55\columnwidth}{\mbox{}\\
\includegraphics[width=0.55\columnwidth]{%
\fig/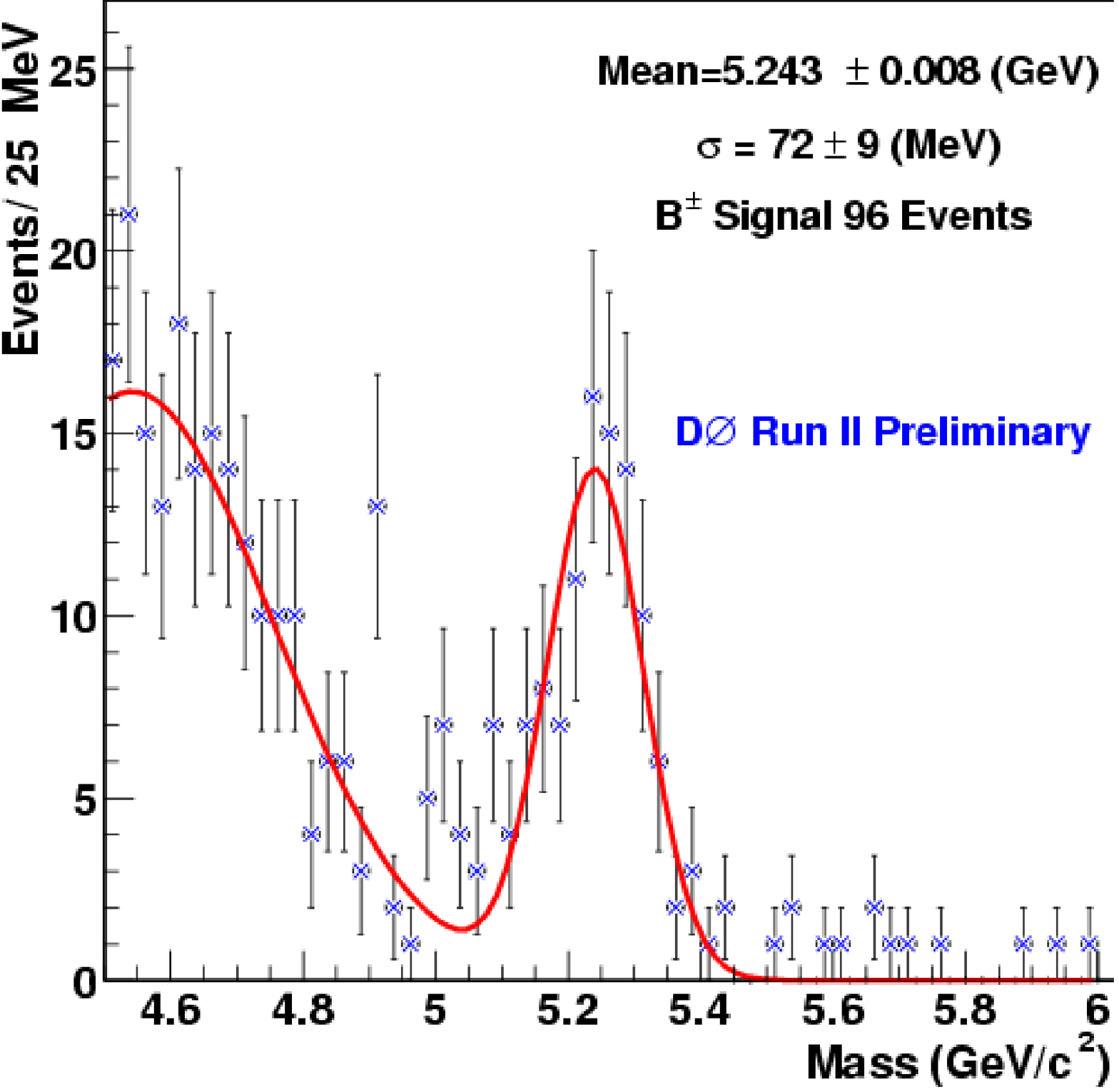}\\
\footnotesize(Mass resolution has improved by factor $\sim 2$ 
since this plot was produced)
}\\
\multicolumn{2}{c}{
\parbox[t]{0.95\columnwidth}{\mbox{}\\
Signal Region:\\
\includegraphics[width=0.99\columnwidth, height=0.66\columnwidth]{%
\fig/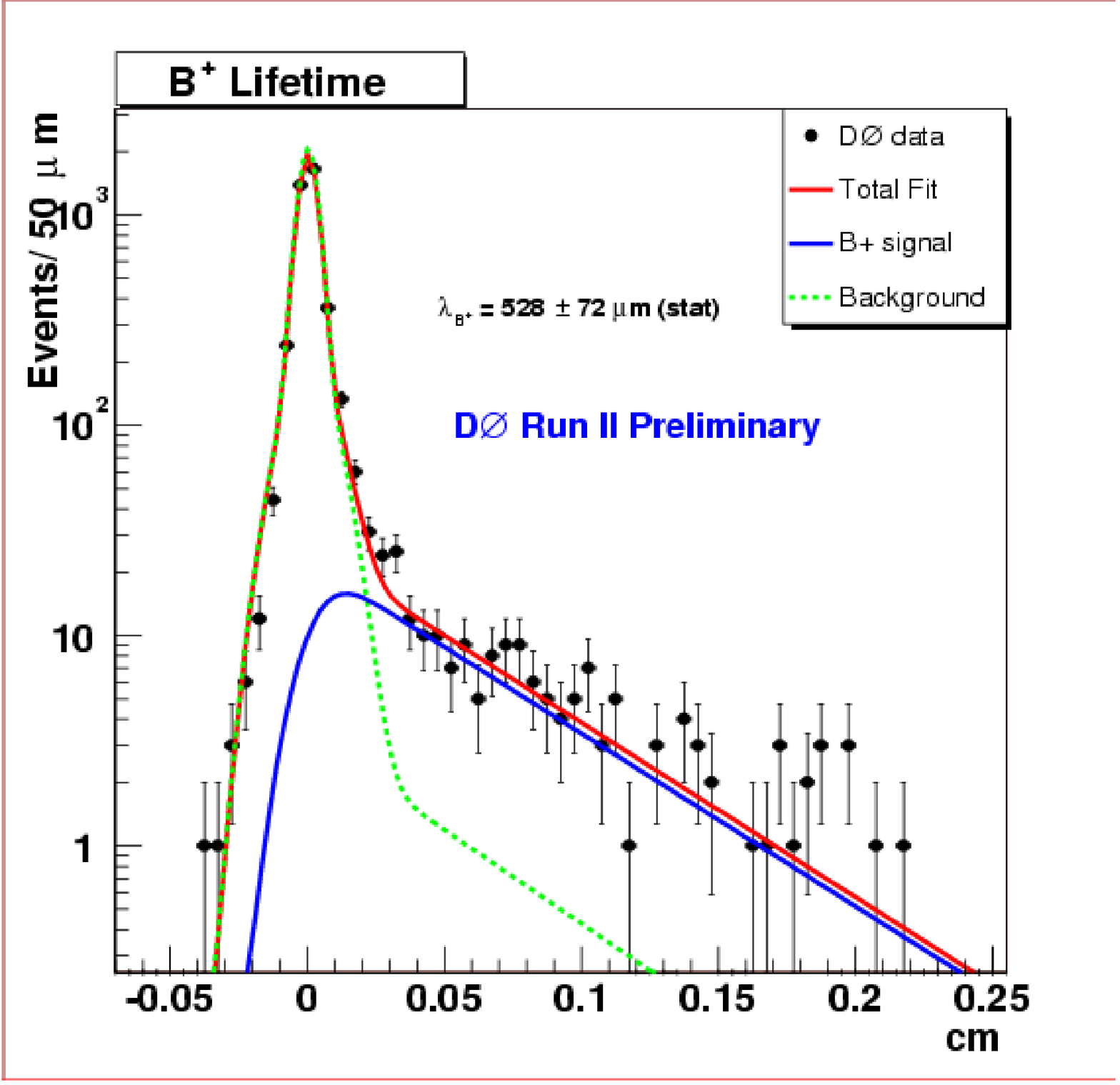}
}
}
\end{tabular}
\caption{Life time fit at \DZERO}
\label{fig:dzeroBplus}
\end{figure}
 \DZERO\ use a somewhat different approach, as illustrated in
 Fig.~\ref{fig:dzeroBplus}, modeling the background using a separate
 fit to the right sideband. The left sideband has a long-lifetime
 component from incompletely reconstructed other B decays. This B
 contamination in the signal region is modeled from Monte Carlo and
 found to be $12\%$.

 The results are given in Table~\ref{tab:BJPsiFull}. The table shows
 that the error on the life time ratios obtained from \prt{B \to
 J/\psi X} decays is about twice that achieved in Run~I, all channels
 combined. By the end of this year, CDF is expected to have collected
 \un{\sim 300}{pb^{-1}}, four times as much as used for the analyses
 presented here, so we can expect CDF to achieve the combined Run~I
 precision using the exclusive channels alone by the end of this year.

\subsubsection{Measurements Using Partially Reconstructed Decays, Without
		IP Trigger}
\paragraph{Inclusive $\tau_B$}
 Since all B hadrons can decay to \prt{J/\psi}, reconstructing
 \prt{J/\psi} vertices allows to find an average B lifetime, where the
 composition of the sample depends on the detector and selection
 criteria. Since the decay is not fully reconstructed, the momentum of
 the B, which is needed to calculate the proper lifetime from
 $\tau(B)=L_{xy} M(B)/( c p_t(B) )$ is unknown. It can however be
 related to the \prt{J/\psi} momentum via
\[
  p_t\left(B\right)=F\left(p_t\left(J/\psi\right)\right) \cdot p_t\left(J/\psi\right)
\]
 where $F\left(p_t\left(J/\psi\right)\right)$ is the mean ratio
 $p_t\left(B\right)/p_t\left(J/\psi\right)$, and the uncertainty on
 $p_t\left(B\right)$ depends on the the spread of that ratio for
 different momenta. Both the mean ratio and its variance are obtained
 from Monte Carlo.
\begin{figure}
 \parbox{\hsize}{
 \rotatebox{90}{\parbox{0.55\columnwidth}{
 {\mbox{}\hfill $\frac{p_t(B)}{p_t(\prt{J/\psi})}$ \hspace{2em}\mbox{}}\\
 \rotatebox{270}{\includegraphics[width=0.88\columnwidth, height=0.55\columnwidth]{\fig/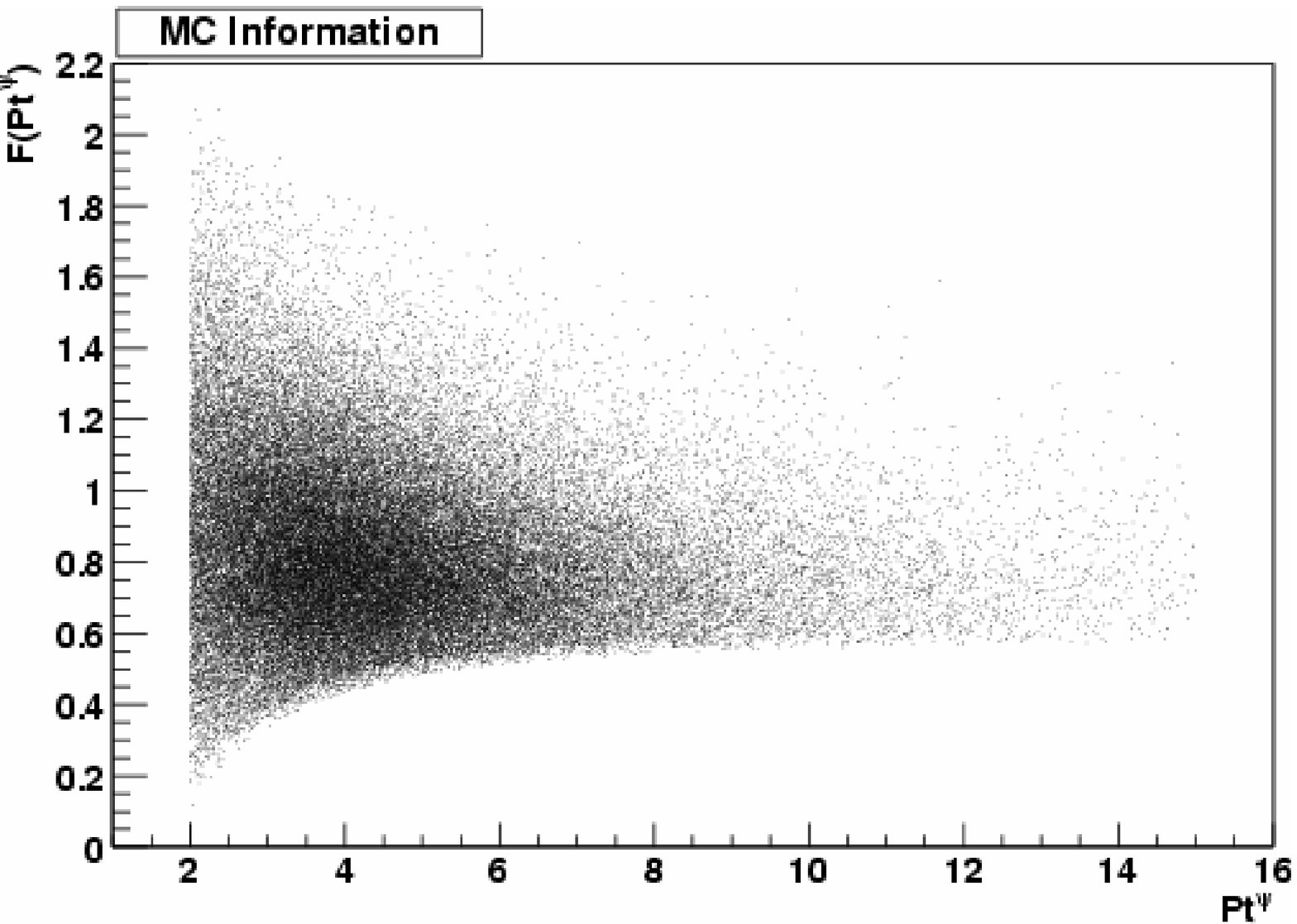}}
 }}\\{\mbox{}\hfill $p_t(\prt{J/\psi})$}}
 \parbox{\columnwidth}{
 \rotatebox{90}{\parbox{0.55\columnwidth}{
 {\mbox{}\hfill $F=\left<\frac{p_t(B)}{p_t(\prt{J/\psi})}\right>$\hspace{2em}\mbox{}}\\
 \rotatebox{270}{\includegraphics[width=0.88\columnwidth, height=0.55\columnwidth]{\fig/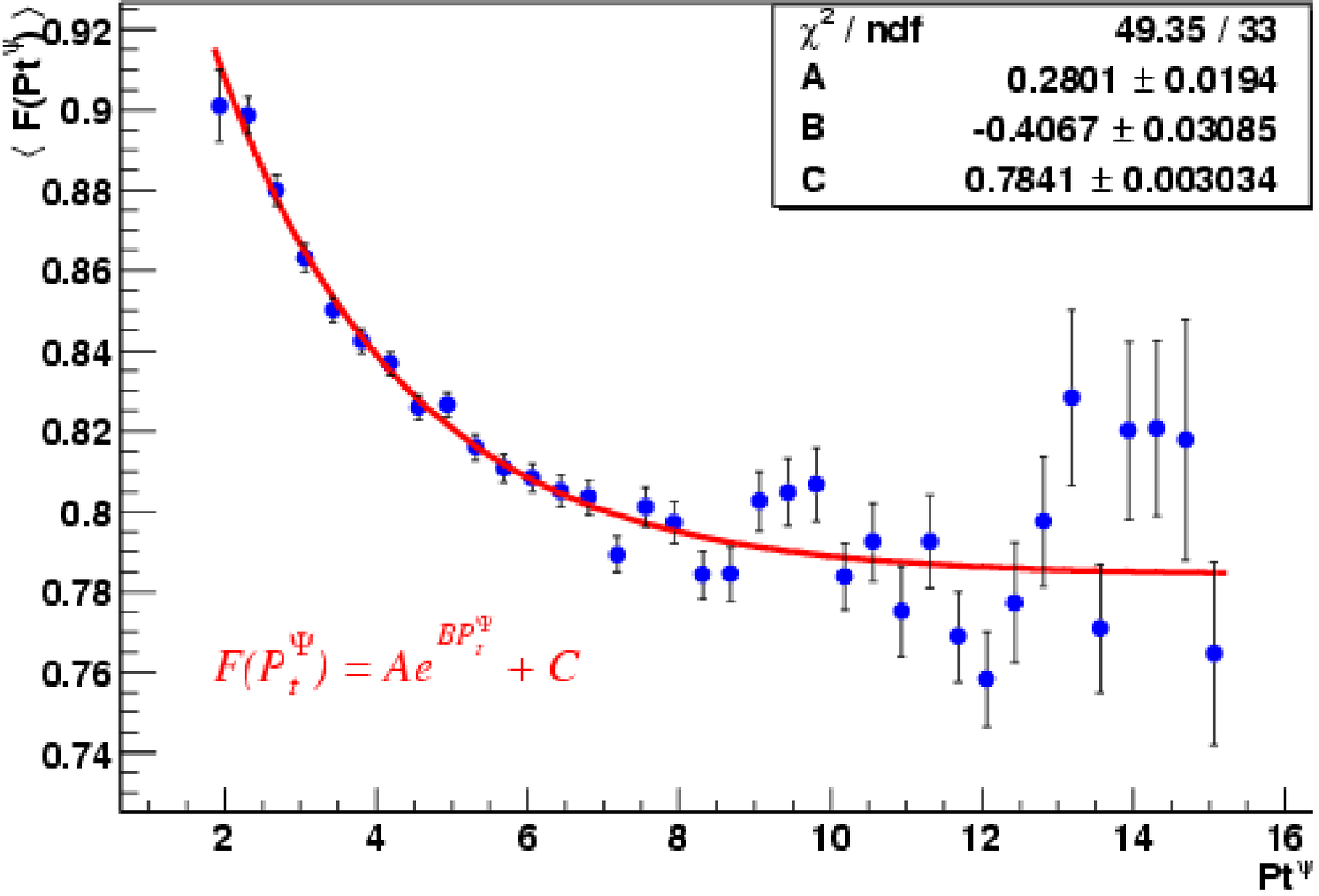}}
 }}\\{\mbox{}\hfill $p_t(\prt{J/\psi})$}}
\caption{The F-factor (\DZERO)}
\label{fig:FFactor}
\end{figure}
 The results of such a Monte Carlo study at \DZERO\ are shown in
 Fig.~\ref{fig:FFactor}.  The measured average B-hadron lifetimes
 are
\begin{itemize}
 \item \DZERO\ (March 03): $\tau_{B} = \un{1.561\pm 0.024 \pm
0.074}{ps}$
 \item CDF (July 02):  $\tau_{B} = \un{1.526\pm 0.034 \pm 0.035}{ps}$
\end{itemize}
 which is consistent with the world average of \un{\tau_{B}=1.573 \pm
 0.007}{ps} \cite{heavyFlavourAveraging}.

\subsubsection{Semileptonic Decays With $\ell +$ IP Trigger}

 CDF is also using \prt{B\to D\ell\nu\;X} and \prt{\Lambda_b \to
 \Lambda_c \ell\nu} decays from the lepton plus displaced track
 trigger for lifetime measurements. The missing $\nu$ momentum is
 accounted for using the same Monte Carlo-based method as in the
 inclusive B lifetime study discussed above. The main challenge is to
 correct the lifetime bias due to the impact parameter cuts in the
 trigger. The acceptance as a function of lifetime
\begin{figure}
\includegraphics[width=\hsize, height=0.6\hsize]{%
\fig/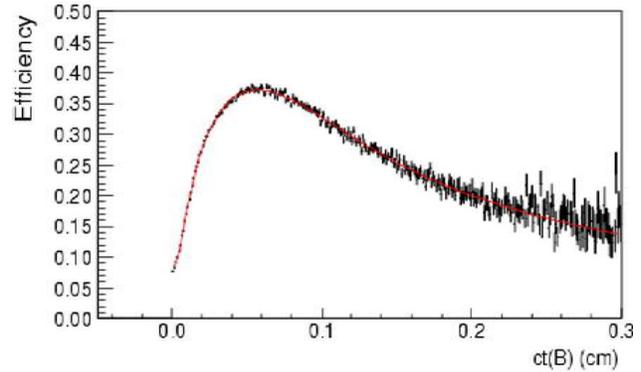}
\caption{SVT acceptance as a function of $c\tau_{B}$ (CDF)}
\label{fig:SVTacceptance}
\end{figure}
 shown in Fig.~\ref{fig:SVTacceptance} is found by a detailed Monte
 Carlo study. A fit to the \prt{B_s \to D_s \mu \nu}
\begin{figure}
\parbox{\hsize}{
\centerline{Sideband \hspace{0.2\hsize} Signal}
\includegraphics[width=\hsize, height=0.5\hsize]{\fig/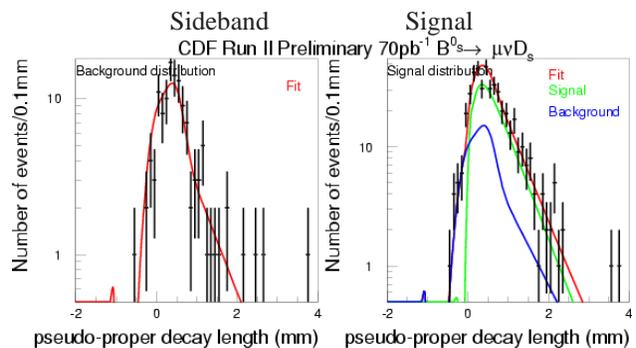}\\
\caption{Fit to lifetime distribution from \prt{B_s \to D_s \mu \nu}}
\label{fig:BsDsMuNuFit}
}
\end{figure}
 life time distribution is shown in Fig.~\ref{fig:BsDsMuNuFit}.
 The statistical precision achieved with the current data
 sample is $\sigma_{\tau_{B^+}}       = \un{\sim 0.05}{ps}$,
           $\sigma_{\tau_{B^0_d}}     = \un{\sim0.06}{ps}$, 
           $\sigma_{\tau_{B_s}}       = \un{\sim0.10}{ps}$,  
           $\sigma_{\tau_{\Lambda_b}} = \un{\sim0.13}{ps}$.
 The full results will be published as soon as the systematic errors
 are fully understood.

\subsubsection{Lifetimes at the Tevatron - Summary \& Prospects}
 The Tevatron is going to provide high statistics samples of all
 B-hadrons, including \prt{B_s, B_c, \Lambda_b}. Preliminary Run~II
 results from fully reconstructed hadronic decays are already
 approaching Run~I precision, higher statistical precision is expected
 soon from the lepton+displaced track sample. The Run~IIa projection
 (MC studies from Dec-01~\cite{BPhysicsAtTev}) for the life time
 ratios are
\begin{itemize}
        \item $\sigma(\tau_{\prt{B_s}}/\tau_{\prt{B^0_d}})$, $\sigma(\tau_{\prt{\Lambda_b}}/\tau_{\prt{B^0_d}}) < 1\%$
\end{itemize}
 which will provide a real test of theory for the \prt{B_s} and,
 pending improved theoretical calculations, for the \prt{\Lambda_b}
 lifetime.

\section{Lifetime Differences}
\subsection{Introduction}
 The width difference between long and short lived CP eigenstates of
 the \prt{B_{s,d}^0-\bar{B}_{s,d}^0} system is predicted to be
\begin{itemize}
 \item $\DGsR \sim \mathcal{O}(10)\%$
 \item $\DGdR \sim \mathcal{O}(1\%)$
\end{itemize}
 \DGsR\ is large enough to be experimentally accessible, soon. The
 width difference is directly proportional to the mass difference,
\[
 \DGs = A \cdot \Delta m_s.
\] 
 where the proportionality constant $A$ is $\sim 3\E{-3}$ in the
 Standard Model, but the value suffers from large hadronic
 uncertainties \cite{Beneke:1998sy}. The mass difference $\Delta m_s$,
 accessible through the oscillation frequency in the \Bso\ system, is
 itself an unknown parameter of great interest, and a major motivation
 for installing the precise vertex detectors at CDF and \DZERO\
 during their upgrades for Run~II. It is interesting to note that a
 large value for $\Delta m_s$, corresponding to fast \Bso\
 oscillations which are more difficult to measure, corresponds to a
 large value for \DGs, which makes it easier to measure.  \DGs\ and
 $\Delta m_s$ are complementary measurements. Given the current limits
 on $\Delta m_s$, a very small value for \DGs\ would be a hint at new
 physics.
\paragraph{Theory Status}
 Recent theoretical predictions for \DGd\ are (the different results
 are obtained using different expansions of Next-To-Leading Order QCD
 corrections:
\begin{itemize}
 \item $\DGdR = \left( 2.6^{+1.2}_{-1.6} \right)\E{-3}$ \cite{Dighe:2001gc}
 \item $\DGdR = \left( 3.0^{+0.9}_{-1.4} \right)\E{-3}$ \cite{Dighe:2001gc} 
 using method in~\cite{BPhysicsAtTev}
\end{itemize}
For \DGs, recent predictions are:
\begin{itemize}
 \item $\DGsR = \left( 8.5 \pm 2.8 \right)\%$~\cite{CKMyellow2002} using method in~\cite{Beneke:1998sy}
 \item $\DGsR = \left( 9.0 \pm 2.8 \right)\%$~\cite{CKMyellow2002} using method in~\cite{Becirevic:2000sj}
\end{itemize}

\subsection{Strategies for Extracting $\Delta\Gamma_s$}
In principle, one could simply fit two exponentials to the lifetime
distribution of \Bso\ decays to mixed CP states. However, since
$\frac{\Delta\Gamma_s}{\Gamma_s} \sim \mathcal{O}(10\%)$ only, this
method would require very high statistics, therefore extra information
is needed to seperate the CP eigenstates. Possible strategies
include~\cite{BPhysicsAtTev}:
\begin{itemize}
\setlength{\itemsep}{1ex plus2ex minus1.1ex}
\item Fit {\color{darkgreen}lifetime} to purely CP-even
        \prt{\Bso\to D_s D_s}. With certain assumptions \prt{\Bso\to
        D^{(*)}_s D^{*}_s} is predicted to be mostly CP even, so that
        these decays could be included in the analysis.
        These assumption would have to be tested
        however, for example with a similar angular analysis as for the
        \prt{\Bso\to J/\psi \phi} case. The result for the CP even
        life time can then be compared to the mean lifetime from
        CP-mixed channels to extract the lifetime difference.
\item Fit 2 {\color{darkgreen}lifetimes} to \prt{\Bso\to J/\psi
        \phi}. This can have 3 angular momentum states, 2 CP even, 1
        CP odd. These can be disantangled by an angular analysis.
\item The {\color{red}B.R.} Method: Assume that the width difference
        is entirely due to CP-even \prt{\Bso\to D^{(*)}_s
        D^{(*)}_s}. In small velocity (Shifman-Voloshin) 
        limit~\cite{BPhysicsAtTev},~\cite{alephDG}:
\begin{displaymath}
             BR(\prt{\Bso\to D^{(*)}_s
             D^{(*)}_s})=\frac{\DGs/\Gs}{1+\frac{1}{2}\DGs/\Gs}
\end{displaymath}
\end{itemize}

\subsection{Current Values for $\Delta\Gamma$}

Recent results for the width differences in the B system are
\begin{itemize}
\setlength{\itemsep}{1ex plus2ex minus1.1ex}
  \item $\DGdR < 0.18$ (95\% CL) (DELPHI, 2002)~\cite{delphiDGd}
  \item $\DGsR < 0.31$ (95\% CL) (combined LEP, CDF, for
         $1/\Gamma_s=\tau(\prt{B_d^0})$, using lifetime method)~\cite{heavyFlavourAveraging}
  \item $\DGsR = 0.26^{+0.30}_{-0.15}$ (ALEPH, from B.R. method)~\cite{alephDG}
\end{itemize}

\subsection{Prospects for $\Delta\Gamma$ at the Tevatron}

 CDF expects the following precisions on \DGs\ by the end of Run~IIa,
 from \un{2}{fb^{-1}}. The projections assume $\DGsR=15\%$ and are
 those given in~\cite{BPhysicsAtTev} in December 2001. They refer to
 the statistical error only.
\begin{itemize} 
\item From \prt{B_s \to J/\psi\phi}:\hfill
 $\sigma_{\mathrm{stat}}(\DGsR)\sim 5\%$
\item \prt{B_s \to D_s D_s} (no \prt{D_s^*}):\hfill
 $\sigma_{\mathrm{stat}}(\DGsR)\sim 6\%$
\item \prt{B_s \to D_s^{(*)} D_s^{(*)}}:\hfill
 $\sigma_{\mathrm{stat}}(\DGsR)\sim 2.5\%$\\ (assume decay 100\% CP even)
\item B.R. method: \hfill $\sigma_{\mathrm{stat}}(\DGsR)\sim 1\%$\\
(model dependent)
\end{itemize}
 Assuming a similar performance for \prt{B_s \to J/\psi\phi} at
 \DZERO\ we arrive at a total statistical uncertainty at the Tevatron
 for \un{2}{fb^{-1}} of $\sigma_{\mathrm{stat}}(\DGsR)\sim 2\%$,
 ignoring the B.R. method. A more conservative estimate of
 $\sigma_{\mathrm{stat}}(\DGsR)\sim 3\%$ is obtained if the assumption
 that \prt{B_s \to D_s^{(*)} D_s^{*}} decays are 100\% CP even is
 dropped and decays involving \prt{D_s^*} are completely ignored.

\section{Conclusion}
\paragraph{Lifetime ratios}
 Since they have started data taking, the B-factories have brought the
 error on $\sigma(\tau_{\prt{B^+}}/\tau_{\prt{B^0_d}})$ down to
 $1.7\%$, so that the experimental accuracy for this ratio is now
 better than that of the HQE prediction. The agreement between
 theory and experiment is very good.  Further improvements on
 $\sigma(\tau_{\prt{B^+}}/\tau_{\prt{B^0_d}})$ can be expected from
 B-factories and Tevatron, soon.

 Large numbers of \Bso\ and \prt{\Lambda_b} are currently being
 produced at the Tevatron. The uncertainty on the lifetime ratios
 $\sigma(\tau_{\prt{B_s}}/\tau_{\prt{B^0_d}})$,
 $\sigma(\tau_{\prt{\Lambda_b}}/\tau_{\prt{B^0_d}})$ is expected to be
 below $1\%$ by the end of Run~IIa. This will provide a real test of
 HQE for \prt{B_s} for which precise predictions exist, while
 improved theoretical values are needed for for \prt{\Lambda_b}.

\paragraph{Lifetime Differences}

 \DGs\ and $\Delta m_s$ are complementary measurements, and both
 parameters combined are sensitive to New Physics
 contributions to \Bso\ mixing. Recent calculations predict 
 $\DGsR = 9 \pm 3 \%$ \cite{CKMyellow2002}.

 From data we get the following limit on the lifetime difference
 in the \prt{B_s} system: $\DGsR < 0.31$~(95\%
 CL)~\cite{heavyFlavourAveraging}.  First steps have been taken
 towards a \DGsR\ measurement at the Tevatron, where $55\pm 9$
 \prt{B_s \to J/\psi\phi} events have been reconstructed at CDF, and
 an average \prt{B_s} lifetime has been extracted from that decay. By
 the end of Run~IIa a measurement of \DGsR\ with a statistical
 uncertainty of $\sim 2\%$ is expected.

\end{document}